\documentclass[10pt,english,american,twocolumn]{article}
\usepackage{bm}%
\usepackage[normalem]{ulem}
\usepackage[colorlinks=true,linkcolor=blue,citecolor=black]{hyperref}%
\usepackage{color}%
\usepackage[T1]{fontenc}
\usepackage[utf8]{inputenc}
\usepackage{algorithm}
\usepackage[font={small}]{caption}
\usepackage{authblk}
\usepackage{float}
\usepackage{mathtools}
\usepackage{amsmath}
\usepackage{amsthm}
\usepackage{amssymb}
\usepackage{graphicx}
\usepackage{graphicx,epstopdf}
\usepackage{algpseudocode}
\usepackage[super]{natbib}
\usepackage{enumerate}
\usepackage{etoolbox}
\usepackage{lmodern}
\usepackage{stfloats}
\usepackage{subfig}

\makeatother

\makeatletter
\def\blfootnote{\xdef\@thefnmark{}\@footnotetext}
\makeatother

\floatstyle{ruled}
\newfloat{algorithm}{tbp}{loa}
\providecommand{\algorithmname}{Algorithm}
\floatname{algorithm}{\protect\algorithmname}

\expandafter\ifx\csname package@font\endcsname\relax\else
 \expandafter\expandafter
 \expandafter\usepackage
 \expandafter\expandafter
 \expandafter{\csname package@font\endcsname}%

\fi
\author[1]{Mattia Serra\thanks{corresponding author: serram@seas.harvard.edu}}
\affil[1]{\footnotesize School of Engineering and Applied Sciences, Harvard University, Cambridge, Massachusetts 02138, USA}%
\author[2]{Pratik Sathe}
\affil[2]{Department of Physics and Astronomy, University of California, Los Angeles, California 90095, USA}
\author[3]{Irina Rypina}
\affil[3]{Physical Oceanography Department, Woods Hole Oceanographic Institution, Woods Hole, MA 02543, USA}
\author[3]{Anthony Kirincich}
\author[4]{Shane D. Ross}
\affil[4]{Aerospace and Ocean Engineering, Virginia Tech, Blacksburg, VA 24061, USA}
\author[5]{Pierre Lermusiaux}
\affil[5]{Mechanical Engineering Dep., Massachusetts Institute of Technology, Cambridge, Massachusetts 02139, USA}
\author[6]{Arthur Allen}
\affil[6]{U.S. Coast Guard Office of Search and Rescue; Washington, DC, 20593, USA}
\author[5]{Thomas Peacock}
\author[7]{George Haller\thanks{corresponding author: georgehaller@ethz.ch}}
\affil[7]{Institute for Mechanical Systems, ETH Zürich, Leonhardstrasse 21, 8092 Zurich, Switzerland}
\title{Search and rescue at sea aided by hidden flow structures}%

\begin{document}
\date{\vspace{-8ex}}
\date{}

\maketitle

\begin{abstract}
\noindent Every year hundreds of people die at sea because of vessel and airplane accidents. A key challenge in reducing the number of these fatalities is to make Search and Rescue (SAR) algorithms more efficient. Here we address this challenge by uncovering hidden TRansient Attracting Profiles (\textit{TRAPs}) in ocean-surface velocity data. Computable from a single velocity-field snapshot, TRAPs act as short-term attractors for all floating objects. In three different ocean field experiments, we show that TRAPs computed from measured as well as modelled velocities attract deployed drifters and manikins emulating people fallen in the water. TRAPs, which remain hidden to prior flow diagnostics, thus provide critical information for hazard responses, such as SAR and oil spill containment, and hence have the potential to save lives and limit environmental disasters.
\end{abstract}
\section{Introduction}
\noindent In 2016, the United Nation Migration Agency recorded over 5000 deaths among people trying to reach Europe by crossing the Mediterranean Sea \cite{DeathAtSeaMediterr2017,DeathAtSeaStatsMSF2017}. This calls for an enhancement of the efficiency of SAR at sea\cite{UNHCR_StrenghtenSARMediterr2018}, which requires improved modeling of drifting objects, as well as optimized search assets allocation  (see\cite{Breivik2013,stone2013search} for reviews). Flow models used in SAR operations combine sea dynamics, weather prediction and in situ observations, such as self-locating datum marker buoys\cite{allen1996performance} deployed from air, which enhance model precision near the last seen location. Even with the advent of high-resolution ocean models and improved weather prediction, however, SAR planning \textcolor{black}{is still based on conventional practices that do not use more recent advances in understanding transport in unsteady flows}. 
 
Current SAR procedures\cite{Kratzke2010} approach uncertainties through Bayesian techniques, turning the modeling exercise into an ensemble integration over all unknown parameters and incorporating unsuccessful searches into locating the next target. This strategy produces probability-distribution maps for the lost object’s location, which, based on a list of assigned search assets, returns search plans, \textcolor{black}{such as planes flying in a regular grid pattern\cite{Kratzke2010}}. The vast uncertain parameter space together with the continuous motion of floating objects driven by unsteady flows, however, leads to error accumulation, ``making SAR planning as much art as science, where rescuers still often rely as much on their hunches as on the output of sophisticated prediction tools''\cite{Breivik2013}.
Furthermore, the convergence of updated probability computations based on a selected prior and unsuccessful searches is usually a slow process, while timing is everything when lives are on the line. 

 \begin{figure*}[h!]
 	\centering
 	\includegraphics[height=.8\columnwidth]{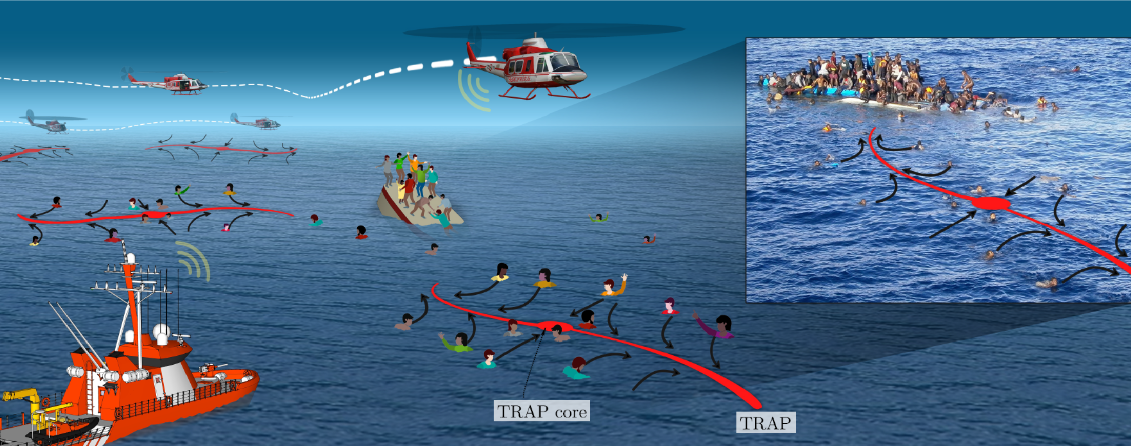}
 	\caption{Sketch of a TRAP-based SAR operation. TRAPs (red curves) emanate from an attracting core (red dot) where their normal attraction (black arrows) is maximal. Different TRAPs provide continuously updated and highly specific search paths. The inset shows a migrant boat that capsized on 12 April 2015 in the Mediterranean Sea along with a schematic TRAP and persons in water. Photo credit: Opielok Offshore Carrier.}	
 	\label{fig:DeathByRescue}
 \end{figure*}

In a SAR scenario, one would ideally have a simply interpretable tool based on key features of the ocean surface dynamics. Such a tool should narrow down the search area by promptly providing the most attracting regions in the flow toward which objects fallen in the water at uncertain locations likely converge. 
This raises the question: How can one rigorously assess short-term variabilities of material transport in fast-changing flows characterized by high uncertainties? 
Here, we address this question using the recently developed concept of Objective Eulerian Coherent Structures (OECSs)\cite{SerraHaller2015} from dynamical systems theory. In our context, attracting OECSs uncover hidden \textit{TRansient Attracting Profiles} (\textit{TRAPs}), revealing the currently strongest regions of accumulation for objects floating on the sea surface. TRAPs are quickly computable as smooth curves from a single snapshot of available modelled or remotely sensed velocity fields, providing highly specific information for optimal search-asset allocation (Fig. \ref{fig:DeathByRescue}). The inset in Fig. \ref{fig:DeathByRescue} shows a migrant boat that capsized on 12 April 2015 in the Mediterranean Sea, along with a schematic TRAP attracting people in the water (PIW). 

We confirm the predictive power of TRAPs in three field experiments emulating SAR situations south of Martha's Vineyard in Massachusetts USA. In the first experiment, we compute TRAPs from a submesoscale ocean surface velocity field reconstructed from remotely sensed High Frequency Radar (HFR) data, and show their decisive influence on surface drifters emulating people that have fallen in water at uncertain locations. In actual SAR operations, however, HFR velocity data is generally not available in real time. We address this challenge in our second and third experiments by computing TRAPs from an ocean model velocity field that assimilates in situ experimental information. We then verify the TRAPs' role in attracting and aligning drifters and manikins, simulating PIW, released in their vicinity through targeted deployments. Our analysis reveals a remarkable robustness under uncertainty for TRAPs: even without accounting for \textcolor{black}{wind-drag or inertial effects due to water-object density difference}--typically uncertain in SAR scenarios--,  the TRAPs invariably attract floating objects in water over two-to-three hours. \textcolor{black}{Such short-time predictions are critically important in SAR.}

\section{Methods}

 Short-term variability in flow features (or coherent structures)  is substantial in unsteady flows. These structures, such as fronts, jets and vortices, continue to receive significant attention in fluid mechanics due to their decisive role in organizing overall transport of material in fluids. Such transport is a fundamentally Lagrangian phenomenon, i.e., best studied by keeping track of the longer-term redistribution of individual tracers released in the flow. In that setting, Lagrangian coherent structures (LCSs) have been efficient predictors of tracer behavior in approximately two-dimensional geophysical flows, such as surface currents in the ocean \cite{LCSHallerAnnRev2015}.
 
Larger-scale models and measurements of environmental flows, however, generally produce Eulerian data, i.e., instantaneous information about the time-varying velocity field governing the motion of tracers. These velocity fields can  be integrated to obtain tracer trajectories, but the result of this integration will generally be sensitive to a number of factors. One such set of factors is the exact release time, release location and length of the observation period. Another major sensitivity factor is errors and uncertainties in the velocity field, which either arise from unavoidable simplifications and approximations in modeling, or from inaccuracies in remote sensing. A third source of sensitivity is the necessarily approximate nature of trajectories generated by numerical integration, due to finite spatial and temporal resolution of the velocity data, as well as to approximations in the numerical integration process. All these factors are significant in predictions for SAR purposes: in fast-changing coastal waters, uncertainties both in the available velocities and in the release location and time are high. This has prompted the use of multiple models, stochastic simulations and probabilistic predictions, all of which require substantial time to be done accurately, even though time runs out quickly in these situations.

\begin{figure*}[h!]
	\centering
	\hfill 
	\subfloat[]{\includegraphics[height=.45\columnwidth]{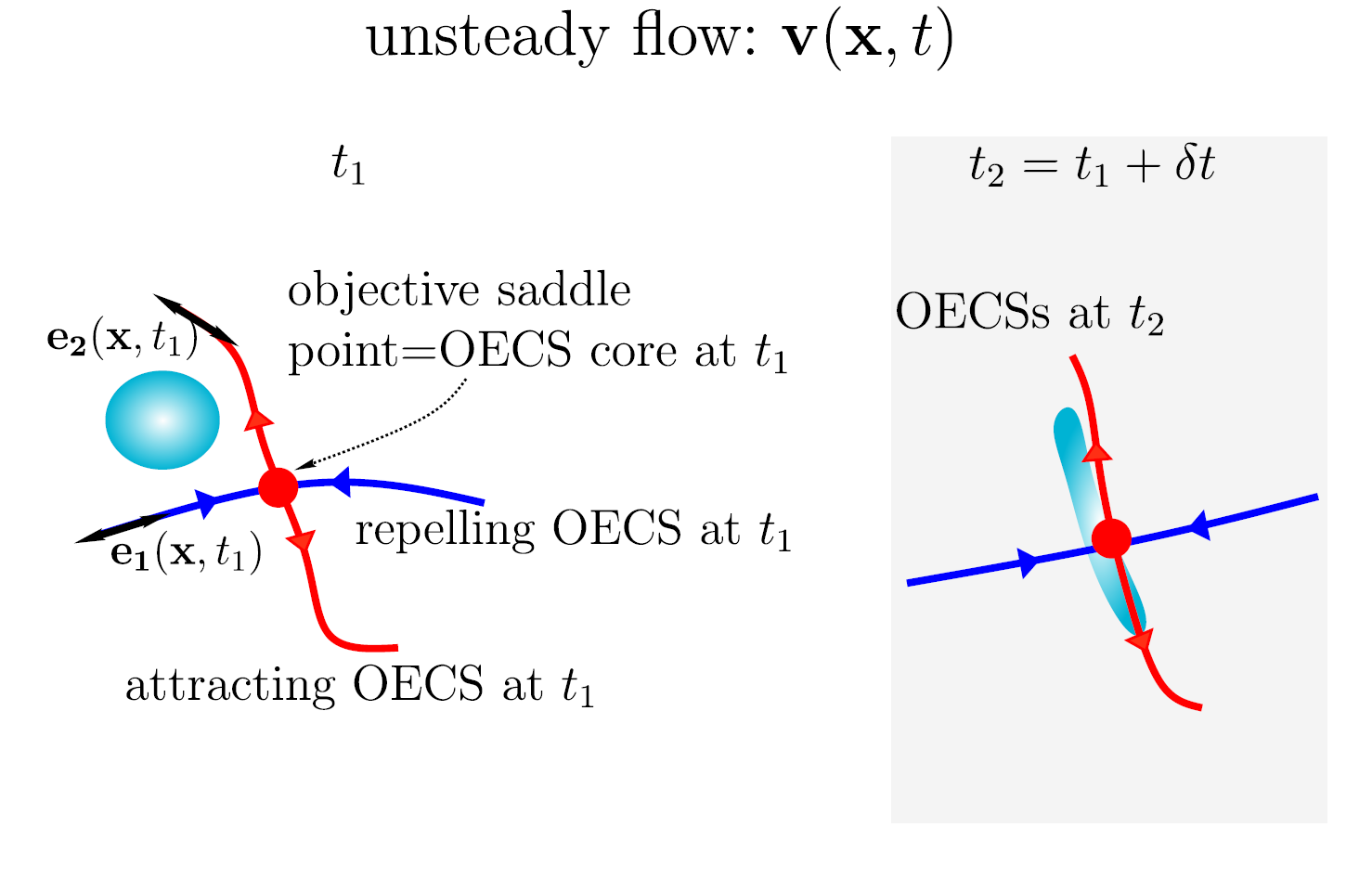}\label{fig:Saddle}}
	\hfill 
	\subfloat[]{\includegraphics[height=.45\columnwidth]{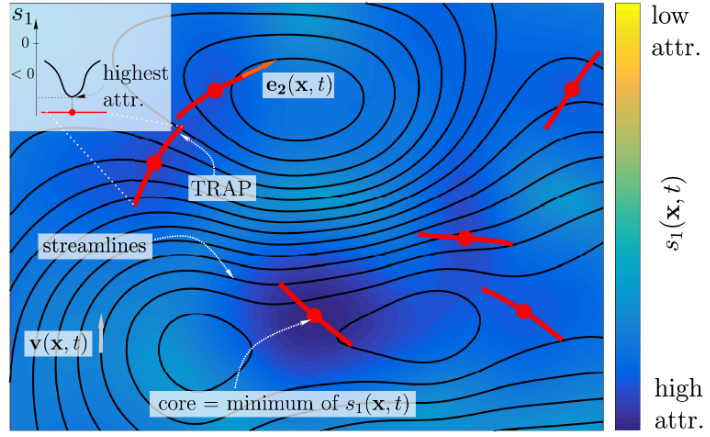}\label{fig:TRAPS_min_s1}}
	\hfill
	\subfloat[]{\includegraphics[height=.45\columnwidth]{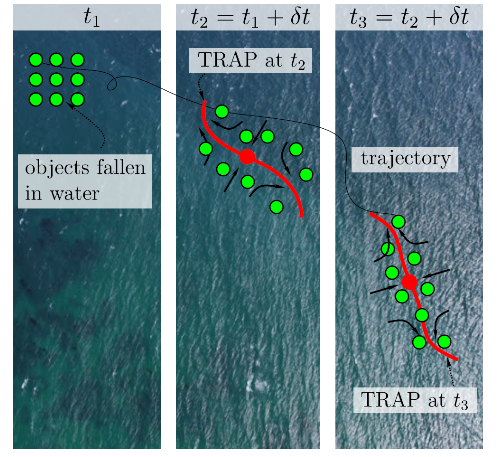}\label{fig:TRAPS}}
	\hfill
	\caption{TRAPs. (a) Deformation of a fluid patch close to an objective saddle point in an unsteady flow. Over short times, a fluid patch aligns with the repelling OECS, and squeezes along the attracting OECS, which both evolve over time. Attracting (Repelling) OECSs are everywhere tangent to the instantaneous $\mathbf{e}_2$ ($\mathbf{e}_1$) fields and their cores are located at minima of $s_1$. (b) Attracting OECSs, i.e. TRAPs, in an unsteady ocean velocity data set derived from satellite altimetry data along with their normal attraction rate $s_1$ encoded in the colorbar. TRAPs are completely hidden to instantaneous streamlines shown in black. (c) Sketch of a TRAP evolving in time and attracting within a few hours floating objects whose uncertain initial locations
		are represented by a square set of green dots.}
	\label{fig:OceanTrap_technical}
\end{figure*}
 
An alternative to these Lagrangian approaches is to find the short-term limits of LCSs purely from Eulerian observations, thereby avoiding all the pitfalls of trajectory integration. These limiting LCSs predict pathways and barriers to short-term material transport until the next batch of updated velocity information becomes available. While simple at first sight, this approach comes with its own challenges, given that  most classic instantaneous Eulerian diagnostics (streamlines, velocity magnitude, velocity gradient, energy, vorticity, helicity, etc) are not objective\cite{TruesdellNoll2004}, i.e., depend on the observer. As such, they cannot possibly be foolproof indicators of material transport, which is a fundamentally frame-independent concept. Indeed, different observers relying on data collected from the coast, from an airplane, from a ship or from a satellite should not come to different conclusions regarding the likely location of materials or people in the water. Yet classic Eulerian quantities would in fact give such different answers (see e.g. Fig. 3a in\cite{LCSHallerAnnRev2015} and Fig. 1 in \cite{SerraHaller2015}). In a SAR situation, this ambiguity is a serious limitation that represents high risk.  

These considerations led to the development of  OECSs\cite{SerraHaller2015}, which are objective (observer-independent) short-term limits of LCSs. Most relevant to our current setting are hyperbolic OECSs in two-dimensional flows, which are the strongest short-term attractors and repellers of material fluid elements. As such, OECSs are extensions of the notions of unstable (and stable)  manifolds of a saddle point in a  steady flow, which attract (and repel, respectively) fluid elements and hence ultimately serve as the theoretical centerpieces of deforming tracer patterns. In unsteady flows and over short times, however, such saddle-type, instantaneous  stagnation points lose their connection with material transport\cite{SerraHaller2015}. Instead,  \textit{objective saddle points} -- the cores of hyperbolic OECSs -- emerge, with associated attracting and repelling  OECSs (Fig. \ref{fig:Saddle}), which, in turn, continuously evolve over time. In our present context,  we will refer to attracting OECSs and objective saddle points as \textit{TRAPs} and \textit{TRAP cores}. 

Unlike stagnation points in steady flows, OECSs cannot be located by inspection of a (frame-dependent) streamline configuration. Instead, consider a planar velocity field $\mathbf{v}(\mathbf{x},t)$, denoting by $\mathbf{e}_2(\mathbf{x},t)$ the dominant (positive) eigenvector and by $s_1(\mathbf{x},t)$ the negative eigenvalue of the rate-of-strain tensor  $\mathbf{S}(\mathbf{x},t)=\tfrac{1}{2}(\mathbf{\nabla v}(\mathbf{x},t) +[\mathbf{\nabla v}(\mathbf{x},t)]^*)$, TRAPs are short segments of curves tangent to $\mathbf{e}_2$ that emanate from local minima of $s_1$\cite{SerraHaller2015} (see Supplementary Information for details). As an illustration,  Fig. \ref{fig:TRAPS_min_s1} shows TRAPs in an unsteady ocean velocity data set derived from AVISO satellite altimetry (see \cite{SerraHaller2015} for a detailed OECSs analysis of this flow). Thus, the $s_1$ scalar field along with the TRAPs provides a skeleton of currently active attracting regions in the flow along with their relative strengths. This in turn gives specific and actionable input for SAR asset allocation, such as high-priority flight paths for discovering people in the water (Fig. \ref{fig:DeathByRescue}).  Remarkably, such pathways remain generally hidden in streamline plots, and can even be perpendicular to streamlines as illustrated in Fig. \ref{fig:TRAPS_min_s1}. Figure \ref{fig:TRAPS} shows that TRAPs evolve over time and attract floating objects whose uncertain initial positions are represented by an array of green dots.

As Eulerian objects, TRAPs are simply computable from a single snapshot of the velocity field $\mathbf{v}(\mathbf{x},t)$. Moreover, velocity fields used in SAR  are generally obtained from models that assimilate environmental data in the proximity to the last known position of a missing person\cite{Kratzke2010,Breivik2013}. This represents a further challenge to Lagrangian prediction methods, as much of their trajectory forecasts tend to leave the domain of reliable velocities and hence have questionable accuracy.  In the Supplementary Information, we illustrate this effect, showing that Lagrangian methods provide only partial coverage when velocities are available over a finite-size domain. A TRAP-based analysis is, therefore, not only faster but provides complete coverage by exploiting all available velocity data.

Finally,  owing to the structural stability of its construction\cite{SerraHaller2015}, TRAPs necessarily persist over short times and are robust to perturbations of the underlying velocity field. In the Supplementary Information we show that the sensitivity of TRAPs to uncertainties is typically lower compared to those of trajectory-based methods. This makes TRAPs a trustworthy now-casting tool for material transport, one that is resilient under uncertainties in initial conditions and other unknown factors, such as the inertia of a drifting object or windage effects.

\section{Results}
Here we show how TRAPs accurately predict short-term attracting regions to which objects fallen in water at uncertain nearby locations converge in ocean field experiments carried out south of Martha’s Vineyard. 
\begin{figure*}[htb!]
	\centering
	\hfill 
	\subfloat[]{\includegraphics[height=.45\columnwidth]{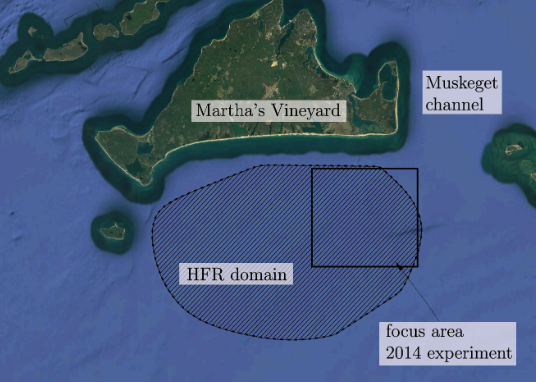}\label{fig:HFRDomain}}
	\hfill 
	\subfloat[]{\includegraphics[height=.45\columnwidth]{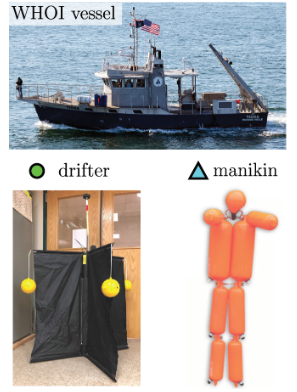}\label{fig:DrifterManikin}}
	\hfill	\subfloat[]{\includegraphics[height=.45\columnwidth]{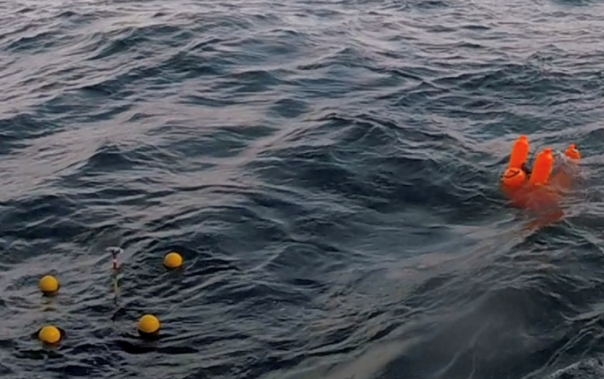}\label{fig:DrifterManikinWater}}
	\hfill
	\hfill
	\caption{Field experiments tools and area of interest. (a) The domain of the ocean field experiments is located south of Martha's Vineyard, where the ocean surface sub-mesoscale velocity, remotely sensed from High-Frequency-Radar (HFR) measurements as described in \cite{Kirincich2012}, is available within the hatched black polygon. The black rectangle represents the area of interest of the 2014 field experiment. (b) Tioga WHOI vessel, CODE surface drifter, whose GPS-tracked position will be marked with a green dot, and OSCAR Water Rescue Training manikins whose GPS-tracked position will be marked with a magenta triangle. (c) Photo illustrating a drifter and a manikin in water during the 2018 experiment. A drone-based video of the 2018 field experiment is available \href{https://www.youtube.com/watch?v=m6jQ9fNK_SU\&feature=youtu.be}{here.}} 	
	\label{fig:HFRandDriftOverview}
\end{figure*}
Figure \ref{fig:HFRandDriftOverview} shows the location of the experiments and the tools we used. In our first experiment, we compute TRAPs from ocean-surface sub-mesoscale velocity derived from High-Frequency-Radar (HFR) measurements available over a uniform 800m$\times$ 800m grid spanning $[-70.7979^\circ, -70.4354^\circ]$ longitude and $[41.0864^\circ, 41.3386^\circ]$ latitude, and in time steps of $30$ minutes. The velocity field is reconstructed from HFR measurements as described in \cite{Kirincich2012}, and is available on a uniform grid within the hatched polygon in Fig. \ref{fig:HFRDomain} (Supplementary Information). 
 
\begin{figure*}[htb!]
	\centering
	\hfill 
	\subfloat[]{\includegraphics[height=.4\columnwidth]{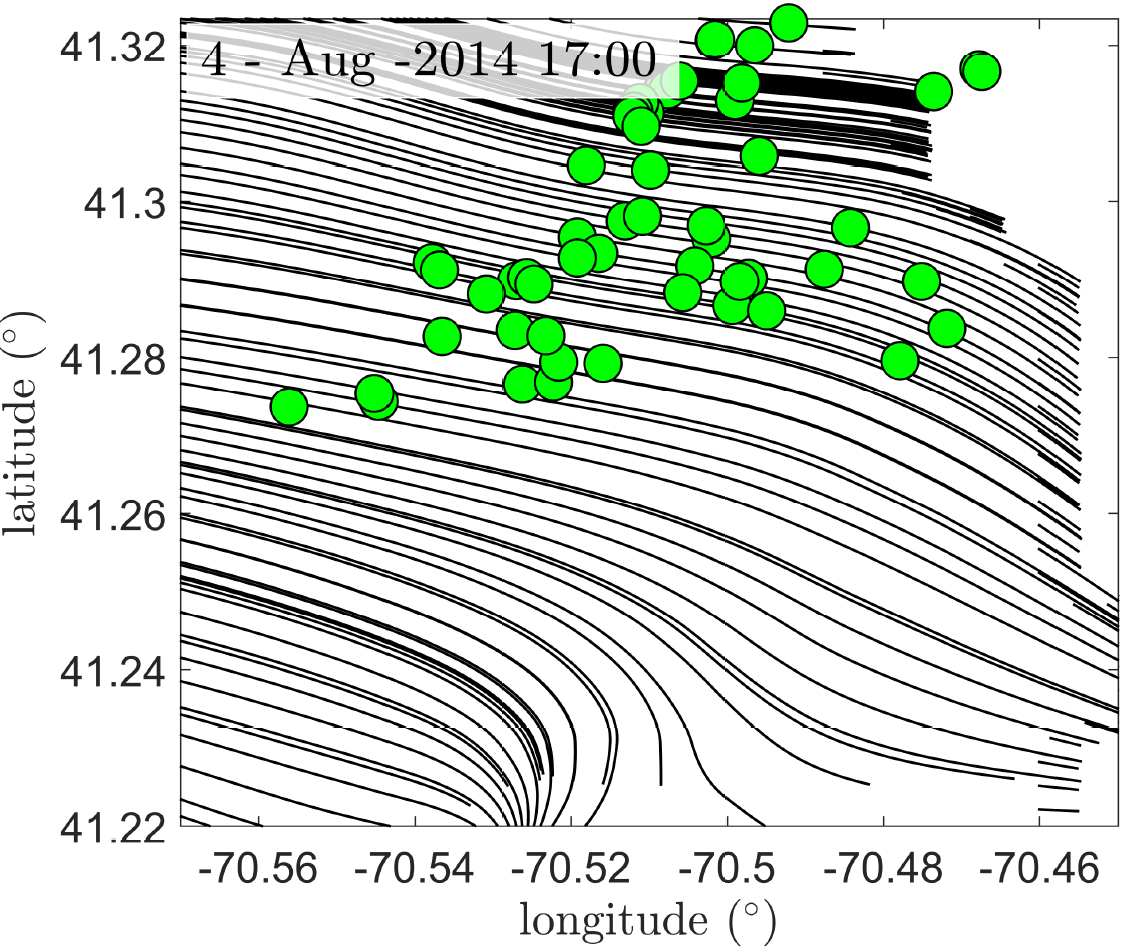}\label{fig:HFR2014_17:00}}
	\hfill 
	\subfloat[]{\includegraphics[height=.4\columnwidth]{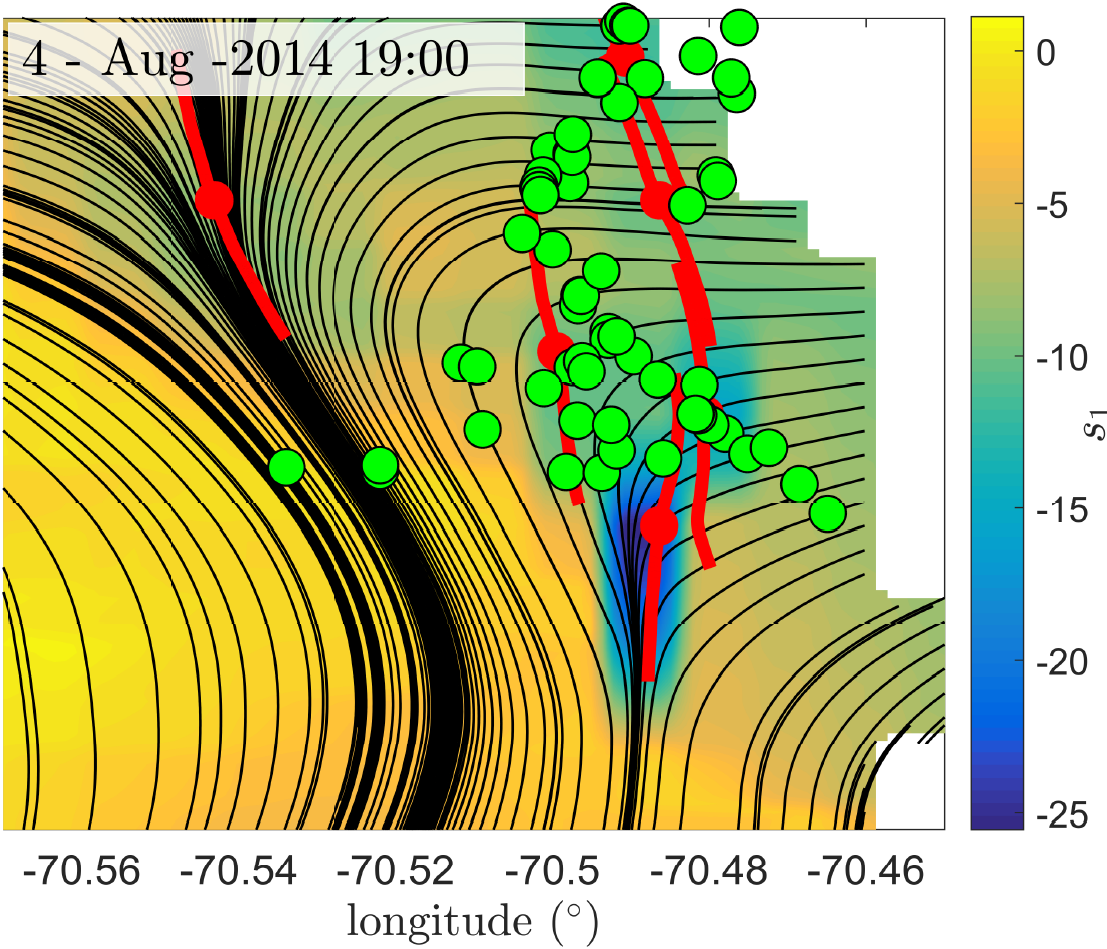}	\label{fig:HFR2014_19:00}}
			\hfill
	\subfloat[]{\includegraphics[height=.4\columnwidth]{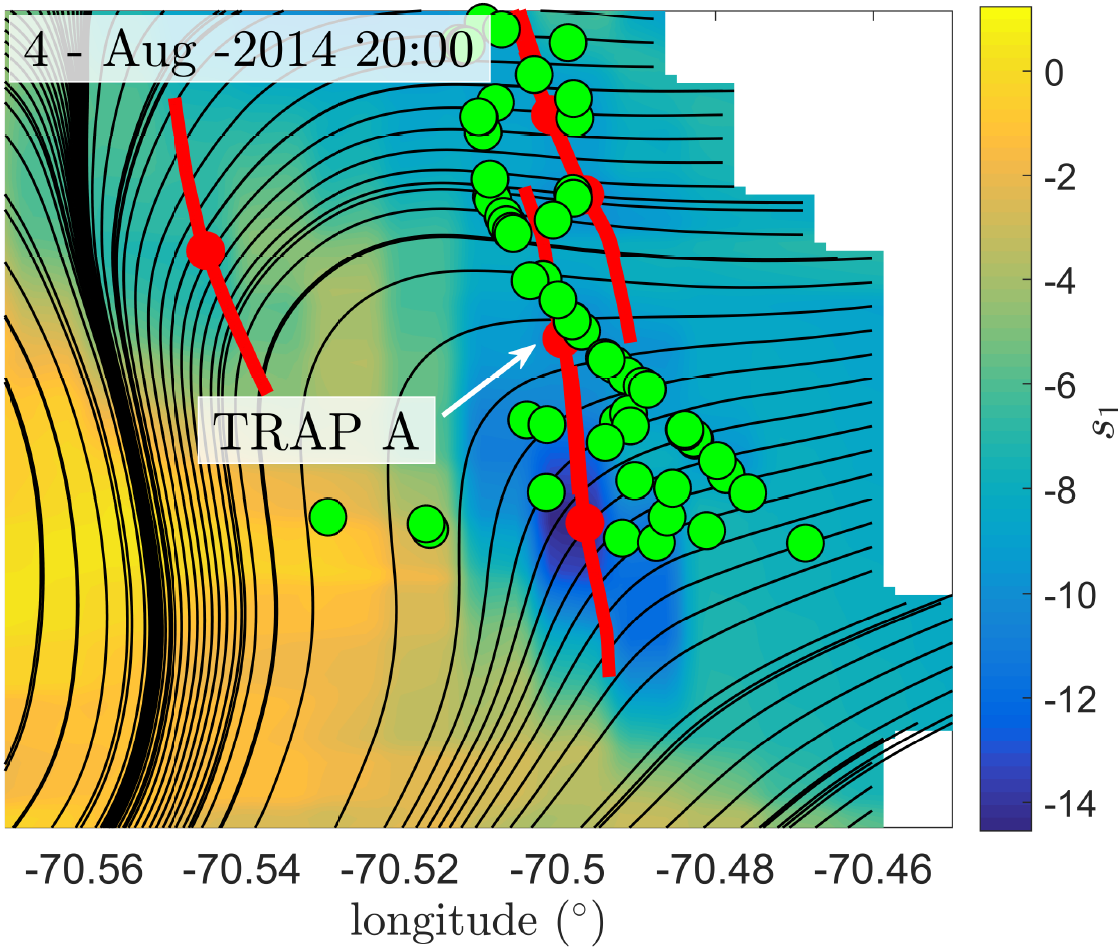}	\label{fig:HFR2014_20:00}}
		\hfill
	\subfloat[]{\includegraphics[height=.4\columnwidth]{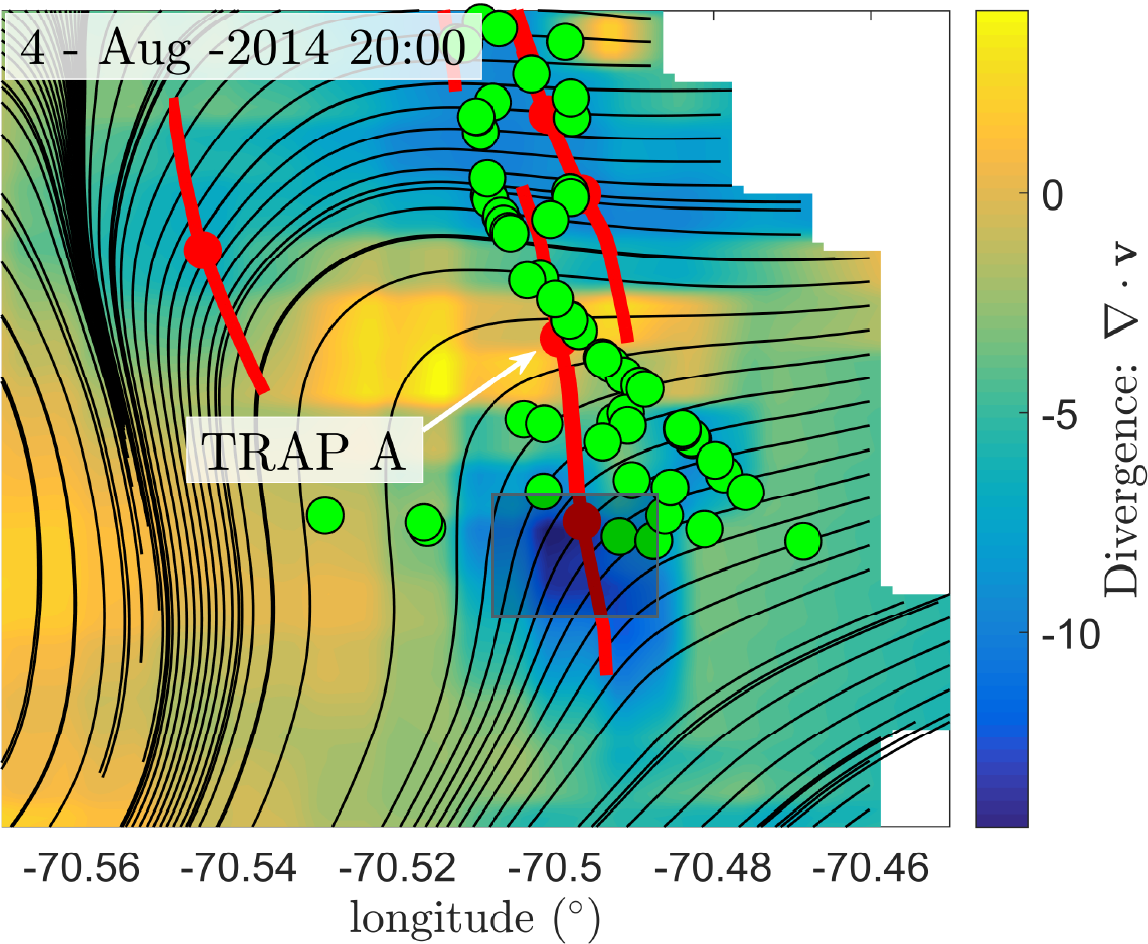}	\label{fig:HFR2014_20:00Div}}	 
	\hfill \hfill
	\caption{2014 experiment. (a) Region bounded by the black rectangle in Fig. \ref{fig:HFRDomain} showing drifter positions (green) at the beginning of our analysis (4th August 2014 at 17:00 EDT), along with the instantaneous streamlines (black) from HFR velocity. (b-c) TRAPs (red curves), whose normal attraction rate $s_1$ (the more negative the more attracting) is encoded in the colorbar, along with instantaneous streamlines and drifters' position, at 19:00 and 20:00 EDT. \textcolor{black}{(d) Same as (c) with the colorbar encoding the divergence field ($\mathbf{\nabla \cdot v}$). TRAP A in panels (c-d) strongly attracts the drifters despite being in a region of positive divergence.} The colorbars unit is 1/day.} 	
	\label{fig:HFRandDrift}
\end{figure*}

To mimic objects fallen in the water, we use 68 Coastal Ocean Dynamics Experiment (CODE) drifters (Supplementary Information and Fig. \ref{fig:HFRandDriftOverview}) whose GPS-tracked locations (green dots) are recorded once every 5min. Drifters of the same design are routinely used by the U.S. Coast Guard in SAR operations.
The starting time of our analysis is the 4th of August 2014 at 17:00 EDT when drifters are located close to the Muskeget channel (Fig. \ref{fig:HFRDomain}). Figure \ref{fig:HFR2014_17:00} shows a zoomed version of the black square inset in Fig. \ref{fig:HFRDomain}, along with drifter positions and the instantaneous streamlines of the HFR velocity. We then compute TRAPs every 30min with the updated velocity field. As expected, we find that the emergence of strong TRAPs at 19:00 (Fig. \ref{fig:HFR2014_19:00}) promptly organize the drifters into one-dimensional structures along TRAPs within two hours (Figs. \ref{fig:HFR2014_19:00}-\ref{fig:HFR2014_20:00}). 

\textcolor{black}{Over longer time scales (approximately a week), drifter accumulation on the ocean surface has been identified with regions of negative divergence\cite{richardson2009drifters,DAsaro2018}. The divergence diagnostic, however, can lead to both false positives and negatives: examples of particle accumulation in regions of zero or positive divergence are given in the Supplementary Information. 
This is precisely the case with TRAP A in Fig. \ref{fig:HFR2014_20:00Div}, which attracts drifters strongly, even though it is located in a region of positive divergence. The negative $s_1$ values along TRAP A (Fig. \ref{fig:HFR2014_20:00}), in contrast, correctly predict its attraction property. Furthermore, regions of negative divergence, irrespective of their validity, tend to be large open sets (Fig. \ref{fig:HFR2014_20:00Div}), as opposed to specific, one-dimensional curves over which we observe drifters clustering.} These results show that TRAPs may be completely hidden in instantaneous streamline and divergence plots, yet predict the short-term fate of passive tracers, as well as inertial objects influenced by windage, such as drifters. Although incorporating inertial, windage and leeway effects could, in principle, provide a better prediction, in a SAR operation the inertia of the target objects is generally unknown\cite{maio2016evaluation} and wind information is unavailable.

Although using HFR velocity would significantly enhance the success of SAR operations\cite{Bellomo2015}, SAR planning is generally based on model velocity data. To account for this, we conducted two more experiments to identify TRAPs from the ocean surface velocity derived from the MIT Multidisciplinary Simulation, Estimation, and Assimilation Systems (MIT-MSEAS)\cite{Haley2010} (Supplementary Information) which assimilates local measurements, similarly to the models used in actual SAR. 

\begin{figure*}[htb!]
	\centering
	\hfill 
	\subfloat[]{\includegraphics[height=.45\columnwidth]{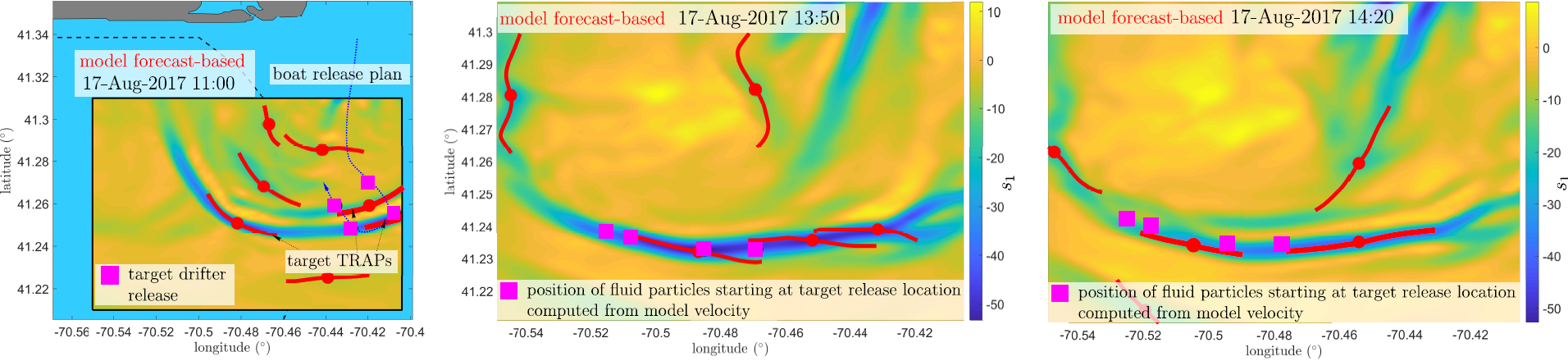}\label{ReleasePlan2017}}
	\hfill 
	\\
	\hfill
	\subfloat[]{\includegraphics[height=.45\columnwidth]{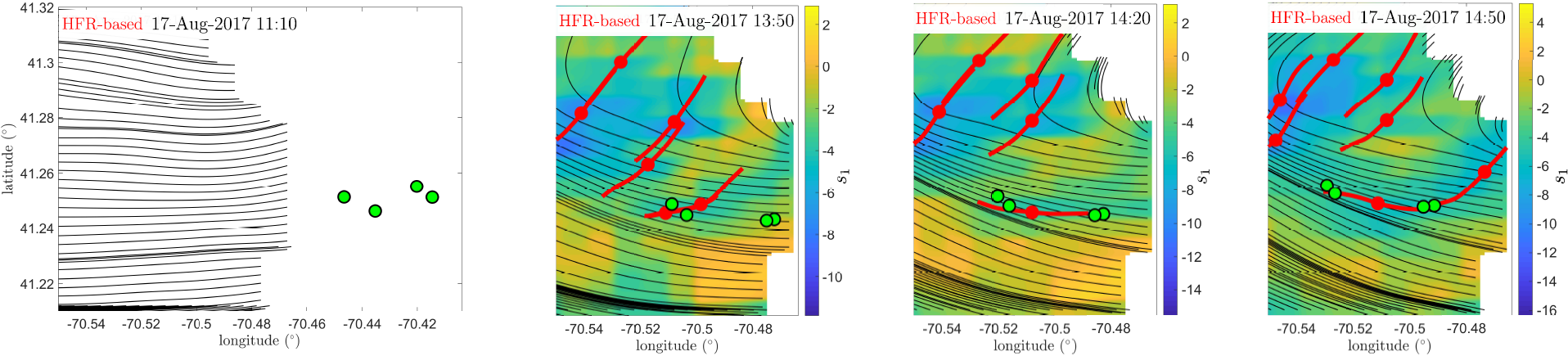}\label{fig:2017HFR}}
	\hfill 
	\caption{2017 experiment. (a) Drifter release plan on the 17th of August 2017 based on the 24h forecast model flow velocity provided on the 16th at 7pm. (Left) Magenta squares denote the target drifter release location at 11:00 am (EDT). (Center-Right) TRAPs from model velocity at later times in the focus region bounded by the black rectangle in Left. Magenta squares are the current position of fluid particles starting at the drifters release location, and computed by integrating the model fluid velocity. (b) Deployed drifters based on (a) along with HFR-based TRAPs. (Left) GPS-tracked drifters position (green dots) at 11:10am along with the flow streamlines computed from the HFR velocity field. The domain corresponds to the focus region in the panels above. The right panels show drifter positions a few hours later, along with the corresponding TRAPs and streamlines computed from the HFR velocity field. The unit of $s_1$ is 1/day.} 	
	\label{fig:2017Experiment}
\end{figure*}

For the experiment performed on the 17th August 2017, we compute TRAPs from the 24h forecast model velocity provided on the 16th August at 7pm. We focus on a region south-east of Martha's Vineyard and identify TRAPs from 11am on 17th August 2017 (Fig. \ref{ReleasePlan2017} Left). The strongest TRAPs are located along a trench of the $s_1(\mathbf{x},t)$ field demarcating a one-dimensional structure containing several TRAPs with strong attraction rates. We note the presence of two parallel trenches from the model. Assuming that the real trench is somewhere in between these two because of modeling uncertainties, we released four drifters north of the lower trench (magenta squares).  The right panels in Fig. \ref{ReleasePlan2017} show later positions of fluid particles obtained by integrating the model velocity field from the target drifter release location, along with the corresponding TRAPs. The figure confirms their attracting property with respect to model data before the float deployment. Based on the release locations in  Fig. \ref{ReleasePlan2017}, Fig. \ref{fig:2017HFR} (left) shows the deployed GPS-tracked drifters position (green dots) at 11:10am within the area of interest bounded by the black rectangle in Fig. \ref{ReleasePlan2017}, along with the streamlines computed from the HFR velocity at the same time. The right panels show later drifter positions along with the TRAPs and the streamlines computed from the HFR velocity field. Although our deployment strategy was purely based on model velocities, the comparison with actual drifter trajectories and TRAPs computed from HFR velocity shows that the model provided reliable estimates of the actual TRAPs. While these TRAPs remained hidden in streamline plots, they nevertheless attracted drifters within three hours.

\begin{figure*}[htb!] 
	\centering
	
	\subfloat[]{\includegraphics[height=.47\columnwidth]{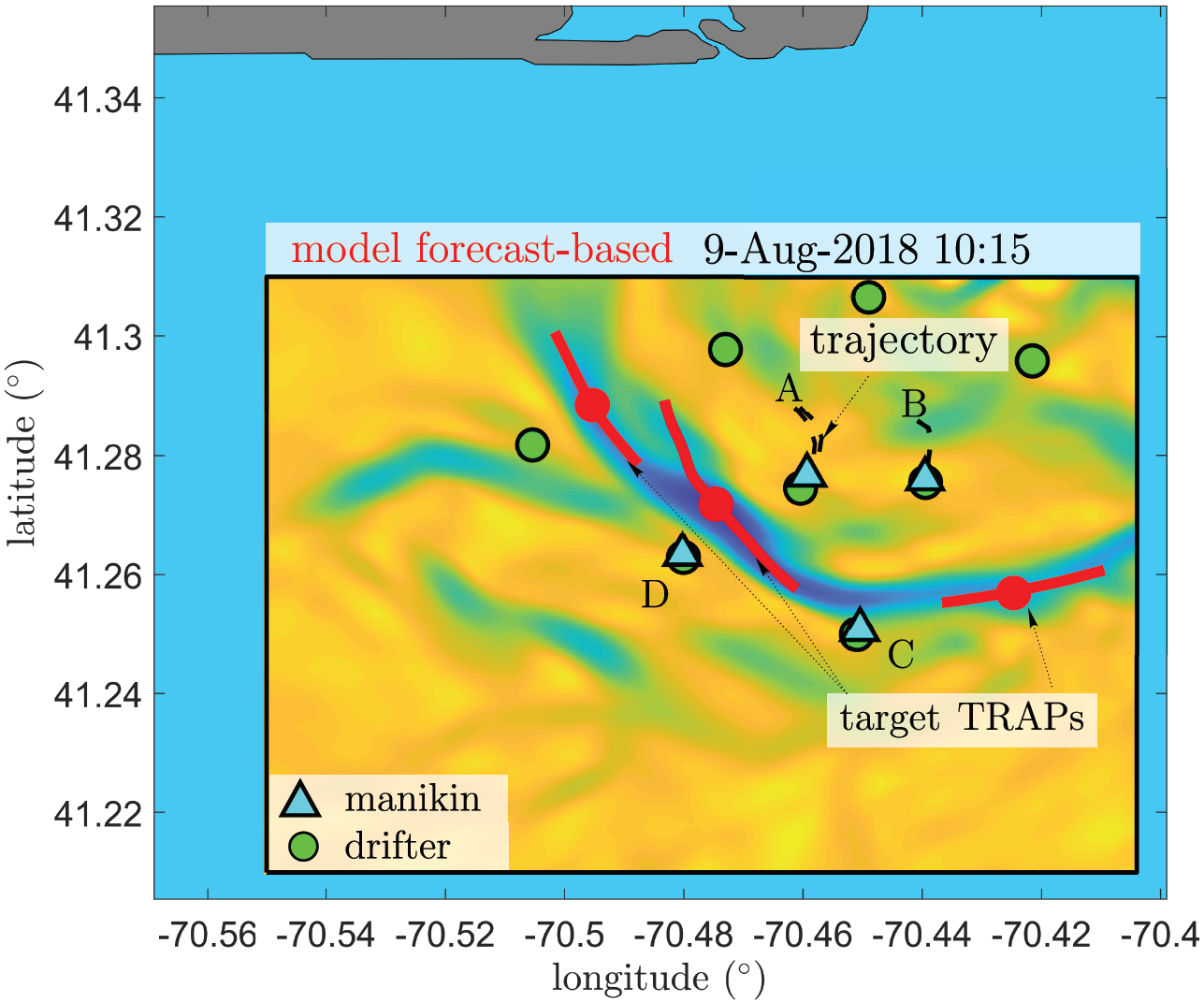}\label{Release2018}}
	\hfill
	\subfloat[]{\includegraphics[height=.47\columnwidth]{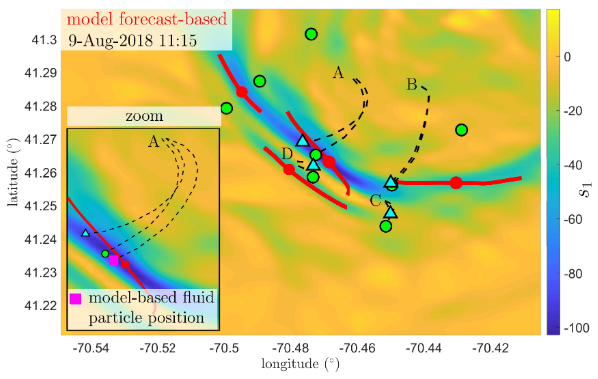}\label{2018Mit_11:15}}
	\hfill
	\subfloat[]{\includegraphics[height=.47\columnwidth]{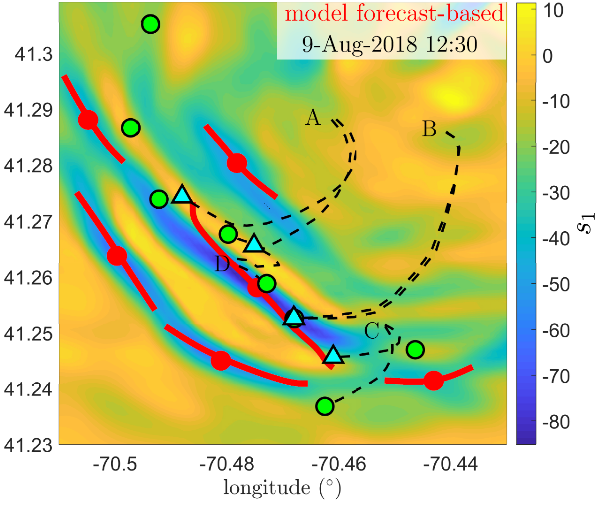}\label{2018Mit_12:30}}
	
	\caption{2018 experiment. (a) Deployed drifters and manikins on the 9th of August 2018 based on TRAPs computed from the 24h forecast model velocity provided on the 8th at 8pm. Green dots and cyan triangles show the GPS-tracked location of CODE drifters and manikins (Fig. \ref{fig:HFRandDriftOverview}) at 10:15am of the 9th August 2018. Dashed lines show object trajectories released at locations A,B,C,D from their deployment to the current time. (b-c) Drifter and manikin positions at later times, along with the corresponding model-based target TRAPs. The inset in (b) shows a zoomed version of the manikin and drifter trajectories released in A, along with the trajectory of a fluid particle (magenta square) obtained by integrating the model velocity from the same initial condition of the drifter and manikin. The unit of $s_1$ is 1/day.} 	
	\label{fig:2018Experiment}
\end{figure*}

In our last experiment, to mimic an even more realistic SAR scenario, we considered a larger set of initially spread-out floating objects consisting of 8 CODE drifters and 4 OSCAR Water Rescue Training manikins manufactured by Emerald Marine Products (Supplementary Information, and Fig. \ref{fig:DrifterManikin}). Using a strategy similar to the 2017 experiment, we designed a deployment for the 9th of August 2018, based on the center forecast model velocity field provided on the 8th of August at 8pm. Figure \ref{Release2018} shows the target model-based TRAPs at 10:15am on the 9th of August 2018, along with all released drifter (green dots) and manikin (cyan triangles) positions. We show only the strongest targeted TRAPs ranked by $s_1$. Dashed curves represent the GPS-tracked trajectories of the deployed objects from their release until 10:15am. In this experiment, we used two WHOI vessels for deployment: one for the release of drifters and manikins at the locations demarcated by A,B,C,D in Fig. \ref{Release2018}, and a second vessel for the remaining drifters. Figures (\ref{2018Mit_11:15}-\ref{2018Mit_12:30}) show the later positions of drifters and manikins along with their trajectories and the recomputed model-based TRAPs\footnote{Because of a relocation of HFR towers in 2018, HFR velocity was not available in the domain shown in Fig. \ref{fig:2018Experiment}.}.
Similar to the previous experiments, both drifters and manikins show a striking alignment with the strongest nearby TRAPs computed from the fluid model velocity within two hours. 

A closer inspection of the deployed drifter and manikin trajectories shows that these two different objects may follow different paths even after short times (less than 2h). This is clearly the case for objects released from locations A,D,C shown in Fig. \ref{fig:2018Experiment}. In the inset of Fig. \ref{2018Mit_11:15}, we show a zoomed version of the drifter and manikin trajectories deployed in A, together with the trajectory of a fluid particle (magenta square) obtained by integrating the model velocity from A. Even though fluid particles, drifters and manikins all follow different trajectories due to inertia, windage and other effects, they invariably converge to the same TRAP, which provides a highly robust attracting skeleton of the underlying flow. In the Supplementary Information, we compare TRAP predictions with trajectory-based ones typically used in SAR. We use nine ensemble velocity field forecasts arising from parametric uncertainty sources 
 (Supplementary Information), and compute the corresponding trajectories using the experimental deployment locations as initial conditions. We find that even though drifter, manikins and ensemble trajectories all differ from each other, they all converge to nearby TRAPs computed from the center-forecast velocity. Using simple mathematical arguments, we also show that TRAPs are intrinsically robust under uncertainties over short times, as opposed to trajectory-based methods, whose sensitivity to uncertainties grow with the largest Lyapunov exponent of the underlying velocity field. Admittedly, TRAPs loose their predictive power over longer time scales because of their instantaneous nature. Shorter time scales, however, are precisely the relevant ones for SAR and hazard response scenarios.

\section{Conclusions}
We have predicted and experimentally verified the existence of TRansient Attracting Profiles (TRAPs), which govern short-term trajectory behavior in chaotic ocean currents characterized by high uncertainties. We expect TRAPs to provide critical information in emergency response situations, such as SAR and oil spill containment, in which operational decisions need to be made quickly about optimal resource allocation. Existing SAR techniques handle uncertain parameters in models of floating objects by averaging several Monte Carlo Simulations and providing probability maps for the objects' location. These maps, however, are not readily interpretable for practical use and can converge slowly due to the underlying chaotic processes. TRAPs and their attraction rates, in contrast, are easily interpretable and highly localized curves which can be computed and updated instantaneously from snapshots of the ocean surface velocity. This eliminates the need for costly trajectory calculations and yields fast input for search-asset allocation. 

We have emulated different SAR scenarios in three ocean field experiments carried out south of Martha’s Vineyard. We computed  TRAPS both from HFR submesoscale ocean surface velocity and from model velocities similar to those available in SAR operations. Our results indicate that TRAPs have significant predictive power in assessing the most likely current locations of objects and people fallen in water at uncertain locations.  
We have specifically found that TRAPs invariably attract nearby floating objects within two-to-three hours, even though they remain hidden to instantaneous streamlines and divergence fields, which also rely on the same Eulerian velocity input. \textcolor{black}{Such a short timing is critical in SAR, as after six hours, the likelihood of rescuing people alive drops significantly}. 
We therefore envision that sea TRAPs will enhance existing SAR techniques, providing critical information to save lives and limit the fall-out from environmental disasters during hazard responses. 
\section*{Acknowledgements}
We are grateful to Margaux Filippi, Michael Allshouse, Javier Gonz\'{a}lez-Rocha, Peter Nolan, Siavash Ameli, Patrick Haley Jr. and the MSEAS team for their contribution in the field experiments. We also acknowledge the NSF Hazard funding grant no. 1520825. S.D.R acknowledges support from NSF grant no. 1821145. M.S. would like to acknowledge support from the Schmidt Science Fellowship \href{https://schmidtsciencefellows.org/}{(https://schmidtsciencefellows.org/)}. G.H. acknowledges support from the Turbulent Superstructures Program of the German National Science Foundation (DFG).
\section*{Author contributions}
M.S. and G.H. designed research; M.S. and P.S. performed research; I.R. provided the drifter data; A.K. provided the HFR velocity data, S.R. provided the manikin data, P.L. provided the model velocity data, A.A. provided knowledge and expertise in SAR; T.P. and I.R. led the field experiments. All authors contributed to the field experiments. M.S., G.H., T.P. and P.S. wrote the paper.
\section*{Competing interests}The authors declare no competing interests. 
\section*{Materials and Correspondence}
All data and codes are available upon request from the corresponding authors.
\setcounter{section}{0}
\onecolumn
\section*{Supplementary Information}
\gdef\thesection{SI.\arabic{section}}
\setcounter{section}{0}
\section{TRAPs in two-dimensional flows}\label{App:CompueSEA TRAPS}
\begin{algorithm}[H]
	\protect \caption{Compute TRAPs}
	\label{algorithm1} \textbf{Input:} A 2-dimensional velocity field
	$\mathbf{v}(\mathbf{x},t)$ 
	\begin{enumerate}
		\item Compute the Jacobian of the velocity field $\mathbf{\nabla} \mathbf{v}$ by numerically differentiating $\mathbf{v}$ with respect to $\mathbf{x}$, and the rate-of-strain tensor $\mathbf{S}(\mathbf{x},t)=\frac{1}{2}\left(\mathbf{\nabla} \mathbf{v}(\mathbf{x},t)+\left[\mathbf{\nabla} \mathbf{v}(\mathbf{x},t)\right]^{*}\right)$
		at the current time $t$ on a grid over the $\mathbf{x} = (x_{1},x_{2}$)
		coordinates, where * denotes matrix transposition. 
		\item Compute the smallest eigenvalue field $s_{1}(\mathbf{x},t)\leq s_{2}(\mathbf{x},t)$ and the
		unit eigenvector field $\mathbf{e_{2}}(\mathbf{x},t)$ of $\mathbf{S}(\mathbf{x},t)$ associated to $s_{2}(\mathbf{x},t)$.
		\item Compute the set $\mathcal{S}_{m}(t)$ of negative local minima of $s_{1}(\mathbf{x},t)$.
		\item Compute TRAPs as solutions of the ODE
		\[
		\begin{cases}
		\mathbf{r}^{\prime}(\tau)=\mathrm{sign}\left\langle \mathbf{e_{2}}(\mathbf{r}(\tau)),\mathbf{r}^{\prime}(\tau-\Delta)\right\rangle \mathbf{e_{2}}(\mathbf{r}(\tau))\\
		\mathbf{r}(0)\in\mathcal{S}_{m},
		\end{cases}
		\]
		where $\tau$ denotes the arclength parameter, $^\prime$ differentiation with respect to $\tau$, and $\Delta$ the arclength increment between two nearby points on the  TRAP. Stop integration when $s_{1}(\mathbf{r}(\tau))>0.3s_{1}(\mathbf{r}(0))$ or $s_{1}(\mathbf{r}(s))\geq 0$.
		
	\end{enumerate}
	\textbf{Output: } TRAPs at time t along with their normal attraction rate field $s_{1}(\mathbf{x},t)$. 
\end{algorithm}
The sign term in step 4 guarantees the local smoothness of the direction field $\mathbf{e_{2}}$, and the termination conditions ensure that the attraction rate of subsets of TRAPs is at least $30\%$ of the core attraction rate, hence exerting a distinguished attraction compared to nearby structures. 
\section{TRAPs and uncertainties}\label{App:Uncertanty}
\begin{figure}[h]
	\hfill
	\subfloat[]{\includegraphics[height=.35\columnwidth]{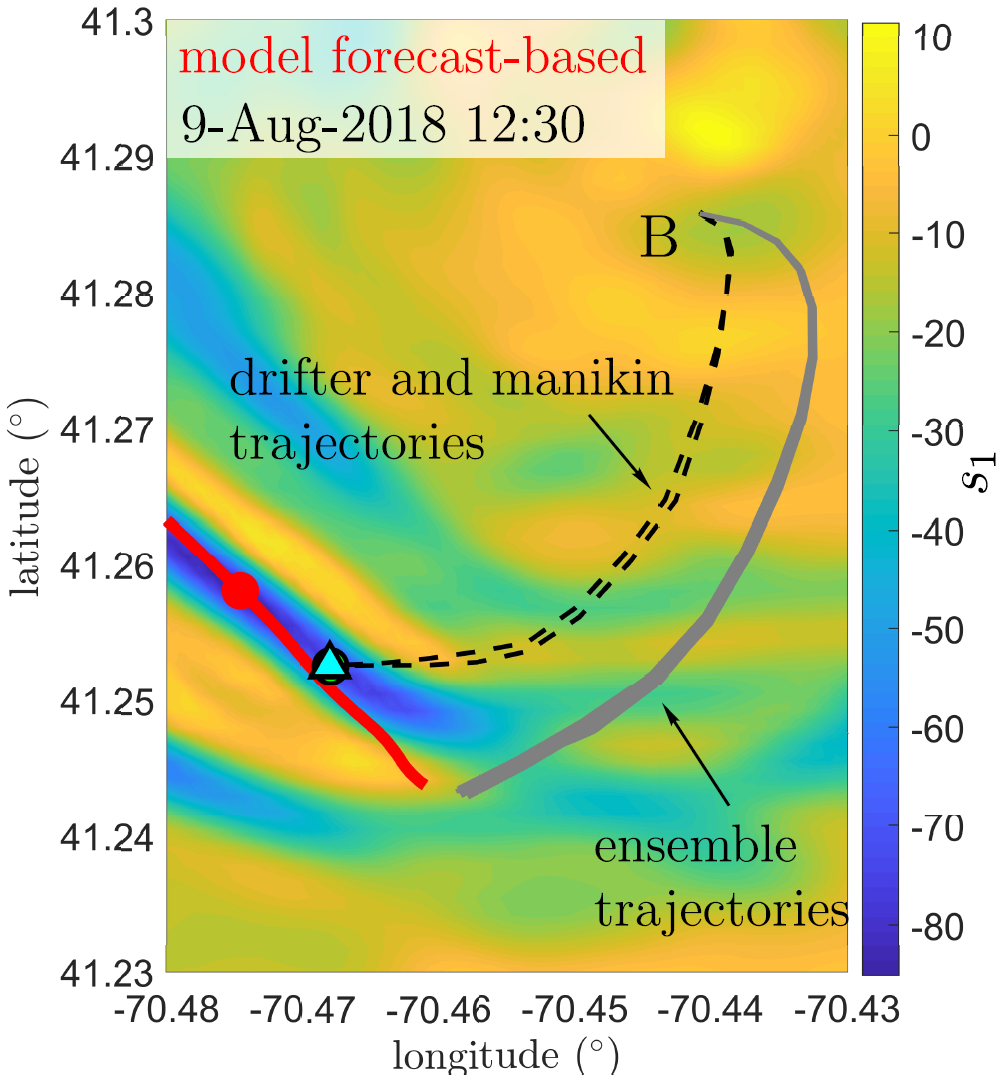}\label{fig:UncertZoom}}
	\hfill
	\subfloat[]{\includegraphics[height=.35\columnwidth]{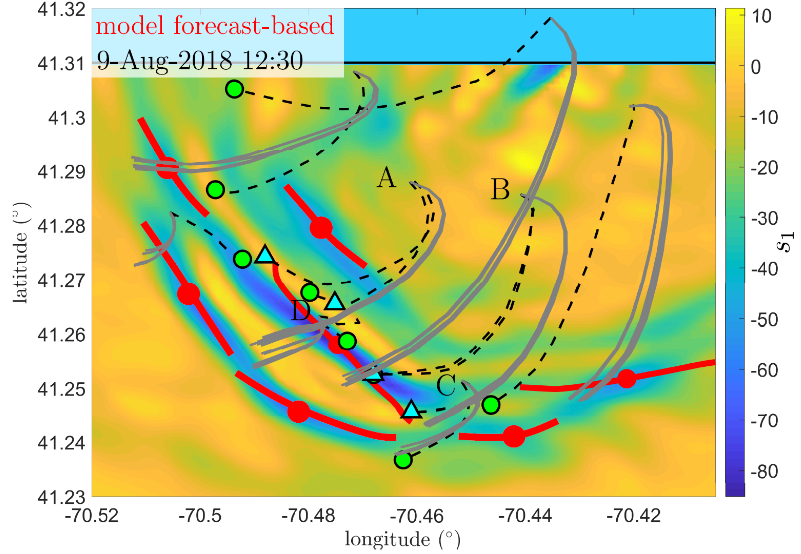}\label{fig:UncertBig}}
	\hfill
		\hfill
	\caption{(a) Drifter and manikin trajectories (dashed lines) released at point B during the field experiment on the 9th of August 2018. Grey lines show trajectories of fluid particles from nine different model based ensemble velocities accounting for uncertainties (see SI, Model Velocity for details). The TRAP is computed from the center-forecast model velocity. (b) Same as (a) for all release locations in the 2018 experiment described in Fig. \ref{fig:2018Experiment}. Despite trajectories from model velocity realizations and real drifters and manikins are visibly different within a few hours, they converge to nearby TRAPs.}
	\label{fig:Uncertanties2018Exp}
\end{figure}

The partial differential equations generating the fluid velocities used in SAR account for a broad set of uncertainties \cite{lermusiaux_JCP2006,lermusiaux_et_al_O2006b}. These affect the initial conditions of all state variables including velocity, as well as external forcing and boundary conditions such as tidal forcing, atmospheric forcing fluxes, lateral boundary conditions, etc. (see SI, Model Velocity for details). With a set of ensemble velocities at hand, typical Lagrangian methods used in SAR compute their corresponding trajectories with the last seen location and time used as initial conditions. Here we consider a total of 9 ensemble forecast velocities modelling part of the above uncertainties during the 2018 field experiment shown in Fig. \ref{fig:2018Experiment}, and compute trajectories of the different ensemble velocities from the drifter and manikin release locations (Fig. \ref{fig:Uncertanties2018Exp}). Figure \ref{fig:UncertZoom} shows the drifter and manikin trajectories, initially released at point B, along with fluid particle trajectoties of the ensemble velocities (gray) and the model-based TRAP computed from the center-forecast velocity. Even within two hours from the release, ensemble trajectories already show visible differences within each other and with the actual drifter and manikin trajectories, yet all converge towards nearby TRAPs. Figure \ref{fig:UncertBig} shows the same as Fig. \ref{fig:UncertZoom} for all the release location of our 2018 experiment.

To gain insights about the predictive power of TRAPs and Lagrangian methods under uncertainties, one can encode the above velocity field uncertainties in the stochastic ordinary differential equation for particle motions $d\textbf{x}(t) = \textbf{v}(\textbf{x}(t),t)dt + \textbf{R}(\textbf{x}(t),t)d\textbf{W}(t)$ where $\textbf{x}(t)$ is the random position vector, $\textbf{v}(\textbf{x}(t),t)$ is the deterministic drift velocity and $\textbf{W}(t)$ is a two-dimensional Weiner process with diffusion matrix $\textbf{R}(\textbf{x}(t),t)$. Then, the uncertain rate-of-strain tensor is $\textbf{S}_u(\textbf{x},t) = \textbf{S}(\textbf{x},t)+\textbf{S}_R(\textbf{x},t)d\textbf{W}(t)$, where $\textbf{S}_R(\textbf{x},t) = \text{sym}(\partial_x\textbf{R}(\textbf{x}(t),t) + \partial_y\textbf{R}(\textbf{x}(t),t))$. In the simplest case of spatially homogeneous uncertainties $\textbf{S}_R(\textbf{x},t) = \mathbf{0}$, hence TRAP predictions remain unaffected while Lagrangian (trajectory based) predictions will have inherent errors that grow both with the largest Lyapunov exponent and $\sqrt{t}$. For TRAPs to be significantly affected by uncertainties,  their spatial inhomogeneities should be comparable to $\textbf{S}(\textbf{x},t)$, which means the model is highly inaccurate. These simple considerations suggest that TRAPs are intrinsically robust predictors under uncertainties over short times. 

\section{TRAPs and velocity field divergence}\label{App:Divergence}
By Liouville’s theorem\cite{ArnoldODE1973}, the infinitesimal phase-space volume of a dynamical system shrinks along a trajectory as long as the trajectory is contained in a domain of negative divergence. Guckenheimer and Holmes\cite{GuckenheimerHolmes1983} conclude that if a steady vector field points everywhere inwards along the boundary of a compact region of negative divergence, then the region contains a nonempty attractor. This criterion, however, will never hold in an unsteady flow on its extended phase space of positions and time. Consequently, there is no applicable, classical dynamical systems
technique to find attractors based on the velocity-field divergence in a general, non-autonomous system. 
This gap has been filled by the variational theory LCS\cite{LCSHallerAnnRev2015} for finite-time flows, and OECS\cite{SerraHaller2015}, as its instantaneous limit.
Indeed, accumulation of particles along lines for longer time intervals is well known to happen along attracting LCS in incompressible velocity fields\cite{Beronetal08b,Olascoaga2013}. 
Our results show that even regions of positive divergence can contain curves that collect drifters/manikins.

\begin{figure}[h]
	\centering
	\includegraphics[height=.24\columnwidth]{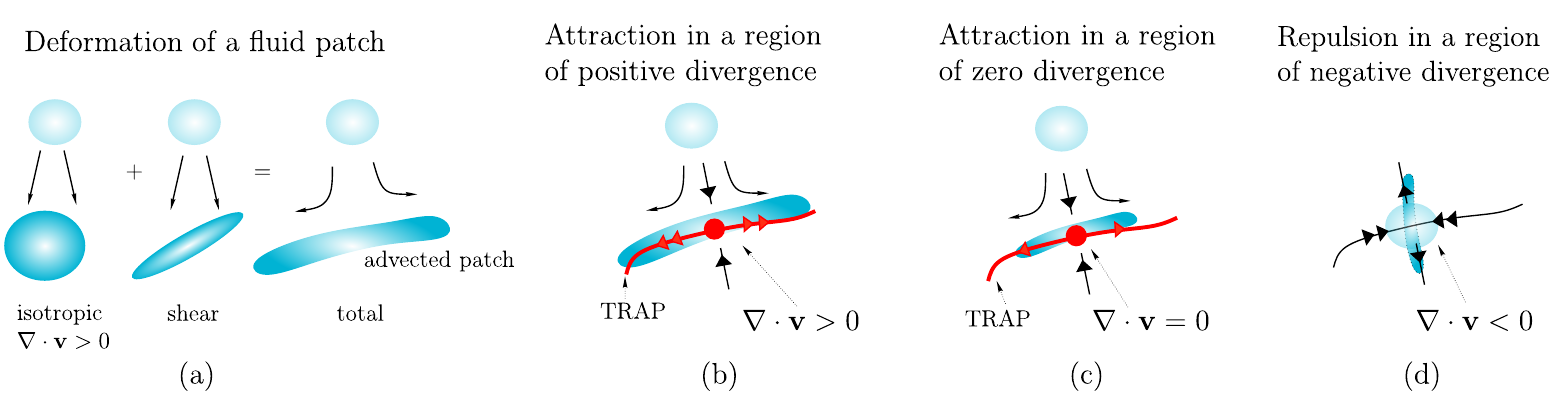}
	\caption{(a) At leading order, the short-term deformation of a fluid patch arises from an isotropic contribution quantified by the divergence of the velocity field ($\mathbf{\nabla \cdot v}$), and an isotropic contribution due to shear. Because of this combination, short-term attraction and clustering can occur in regions of positive divergence (b) or zero divergence (c), invariably captured by TRAPs (red). (d) Tracers escape a region of negative divergence.}
	\label{fig:DivergeneAndSeaTraps}
\end{figure}

The attraction of TRAPs arises from the combination of isotropic and anisotropic deformations (Figure \ref{fig:DivergeneAndSeaTraps}(a)). The isotropic component is due to the instantaneous divergence of the velocity field ($\mathbf{\nabla \cdot v}$), while the anisotropic one to shear, both of which are encoded in $\mathbf{S}$, and thus in the definition of TRAPs.
Figure \ref{fig:DivergeneAndSeaTraps} (b-c) illustrate short-term clustering regions, correctly predicted by TRAPs, where the local divergence is positive or zero. Panel (b) is similar to Fig. 4c, panel (c) to Fig. 2b or to the findings in Ref. \cite{SerraHaller2015}. Figure \ref{fig:DivergeneAndSeaTraps} (d) shows that tracers can escape a region of negative divergence. 
\section{Datasets}\label{App:Dataset}
\subsection{High Frequency Radar Velocity Field}\label{App:HFR Velocity}
The WHOI high frequency radar system (HFR), as operated during the 2014-2017 experiments, consisted of 3 land-based sites spaced at $~$10km intervals along the south side of Martha's Vineyard, MA. These 25MHz systems were run using a combination of 350kHz transmit bandwidth and low transmit power (10W max) which allowed all systems to achieve resolutions of 429m and ranges of 30km  (see \cite{kirincich2016remote} for details). Doppler spectra received from each system were processed using advanced methods\cite{Kirincich2012} into radial velocity estimates every 15min based on a 24min averaging window. Radial velocity estimates were quality controlled before inclusion into the vector velocity estimates using standard time-series QC techniques. These data were combined into vector velocities on a uniform 800m resolution grid, given in latitude and longitude coordinates, using a unique weighted least squares technique that employed non-velocity based signal quality metrics to weight the data to increase the accuracy of the final product. Two successive estimates of the 15min radials are used to estimate the vector (east and north) velocities on a 1/2h time interval centered on the hour. The spatial extent of the vector velocities was limited by theoretical Geometrical Dilution of Precision (GDOP) values less than 1.75.  An error estimate for the east, north, and the norm of the vector velocity components is given.  
This estimate uses the radial velocity error estimates (the weighted standard deviation of the individual HF radar radial returns found within each 5 degree azimuthal bin average) in a standard (numerical recipes) vector error calculation.

\begin{figure}[h]
	\hfill
	\subfloat[]{\includegraphics[height=.33\columnwidth]{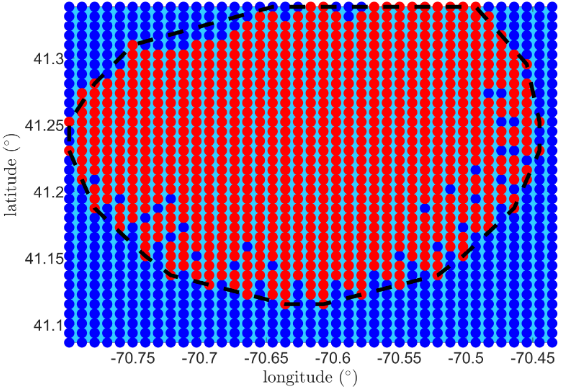}\label{fig:velo_avail_init}}
	\hfill
	\subfloat[]{\includegraphics[height=.33\columnwidth]{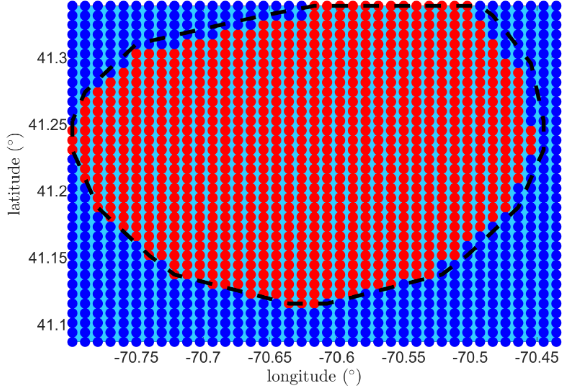}\label{fig:velo_avail_fin}}
	\hfill{}
	\caption{Latitude-longitude grid points where the velocity values are available at a particular time are shown in red. Even within the region of interest bounded by the dashed line, there are blue points where the velocity field is not available due to occasional measurement deficiency. The dashed curve is the convex hull enclosing all the red points. Velocity field grid points (a) before and (b) after interpolation.}
	\label{fig:interpolation_scheme}
\end{figure}

Because of occasional measurement deficiency, there are grid points at which the velocity is not available within the region of interest (blue dots inside the dashed curve in Fig. \ref{fig:velo_avail_init}). To overcome this limitation, we devise a simple interpolation scheme by which we can obtain velocity everywhere within a well-defined boundary. Specifically, we first compute the boundary of this region (the dashed curve in Fig. \ref{fig:interpolation_scheme}), using the Delaunay triangulation function in MATLAB. For each time instance at which the velocity field data is available, we obtain an estimate of the velocity at the blue points inside the boundary using a linear scattered interpolation scheme (see Fig. \ref{fig:velo_avail_fin}). Once the velocity field is available within the black boundary, we convert it from its original units of $m s^{-1}$ to degrees per day. 

Finally, we note that HFR velocities are computed by averaging the raw radial velocity estimates with a 800m window radius at each grid point, spaced 800m apart from each other\cite{kirincich2016remote}. To yield an accurate computation of TRAPs from HFR velocities consistent with the way the data are processed, we smooth $\mathbf{\nabla} \mathbf{v}(\mathbf{x},t)$ with a spatial average filter whose width corresponds to two grid sizes (1600m), as in Ref.\cite{Kirincich2016}. 
\subsection{Finite-size domain}\label{App:finite size}
\begin{figure}[h!]
	\hfill
	\subfloat[]{\includegraphics[height=.31\columnwidth]{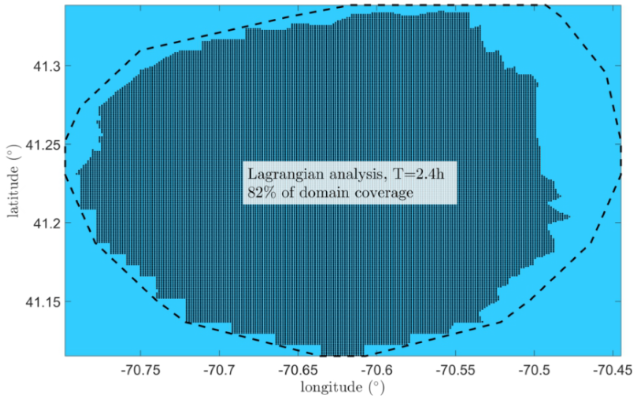}\label{fig:coverage2.4h}}
	\hfill
	\subfloat[]{\includegraphics[height=.31\columnwidth]{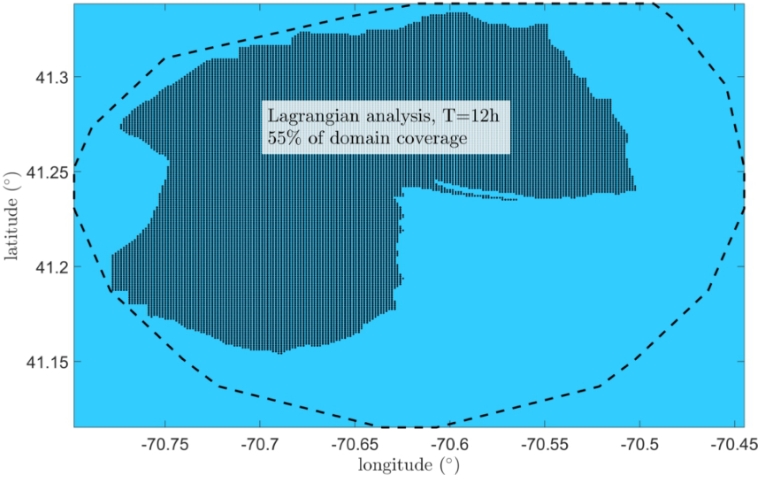}\label{fig:coverage12h}}
	\hfill{}
	\caption{Integrating the HFR velocity field starting at 4th of August 2014 at 8pm, we compute trajectories of fluid particles from a dense set of initial conditions covering the entire domain bounded by the dashed curve.  (a) After 2.4h, 82\% of fluid particles (black points) are within the finite-size domain. (a) After 12h, only 55\% of fluid particles are within the finite-size domain.}
	\label{fig:LagrangianCoverage}
\end{figure}
When the velocity field is available over a finite-sized domain, as the case of HFR velocity, Lagrangian Coherent Structures (LCSs) methods\cite{LCSHallerAnnRev2015,hadjighasem2017critical} invariably provide incomplete coverage 
because of particles leaving the domain. Even if the velocity field is derived from models, they typically assimilate in situ measurements\cite{allen1996performance} to enhance model predictions in a specific region of interest, making the resulting velocity field more accurate in a finite-size domain. 
Assuming that such a region is the domain over which HFR velocity is available, Fig. \ref{fig:LagrangianCoverage} shows the coverage reduction of LCSs methods for two different integration times. We compute fluid trajectories by integrating the HFR velocity field starting on the 4th of August 2014 at 8pm, with a dense set of initial conditions covering the entire domain bounded by the dashed curve. After 2.4h (12h) any Lagrangian method can provide information only on 82\% (55\%) of the region where the velocity field is available. Eulerian methods, instead, provide full coverage of the domain because they do not rely on particle trajectories.

\subsection{Model velocity}\label{App:Model Velocity}

We use the MIT Multidisciplinary Simulation, Estimation, and Assimilation Systems (MIT-MSEAS)\cite{Haley2010,haley_et_al_OM2015} Primitive-Equation (PE) ocean modeling system to compute ocean surface velocity forecasts in the Nantucket and Martha's Vineyard coastal region during August 2017 and 2018. The modeling system is set-up in an implicit 2-way nesting configuration, and provided forecasts of the ocean state variable fields (three-dimensional velocity, temperature, salinity, and sea surface height) every hour with a spatial resolution of 200m in the Martha's Vineyard domain and of 600m in the larger Shelf domain. The ocean forecasts are initialized using historical and synoptic ocean CTD data from the National Marine Fisheries Service (NMFS) and the Martha's Vineyard Coastal Observatory (MVCO), SST images from the Johns Hopkins University's Applied Physics Lab (JHU APL), and other data from available sources. These ocean simulations are forced by atmospheric flux fields forecast by the National Centers for Environmental Prediction (NCEP) and tidal forcing from TPXO8, but adapted to the high-resolution bathymetry and coastlines\cite{logutov_lermusiaux_OM2008}. \textcolor{black}{Finally, in situ CTD measurement have been assimilated into the modeling system at the start of the experiment within the first few days of August.}

The deterministic 2-way nesting ocean forecast initialized from the estimated ocean state conditions at a particular time is referred to as the central forecast. The ensemble forecasts were initialized using Error Subspace Statistical Estimation procedures \cite{lermusiaux_JAOT2002}. The forecasts within the ensemble were commonly initialized from perturbed initial conditions of all state variables (temperature, salinity, velocity, sea surface height) and forced by perturbed tidal forcing, atmospheric forcing fluxes and lateral boundary conditions. These initial, forcing and boundary perturbations are created so as to represent the expected uncertainties in each of these quantities, respecting ocean physics and in accord with the few observed data misfits. Each ensemble members were 2-way nested in two domains and required respecting domain embedding conditions. Finally, parameter uncertainties (bottom drag, mixing coefficients, etc.) were also modeled by perturbing the values of parameters for each ensemble forecast. 

Ensemble forecasts were issued twice daily during the 2 weeks of the 2017 and 2018 experiment (see \url{http://mseas.mit.edu/Sea_exercises/NSF_ALPHA/2018/}), with the number of ensemble forecasts issued varying between 7 and 100 depending on the number of computing units available and on computational power. The nine 2018 MSEAS ensemble forecasts utilized for the present study correspond to parametric uncertainties only, representing uncertainties in the surface wind mixing, tidal mixing and tidal bottom friction.

\subsection{Drifters}\label{App:Drifters}
The drifters used in our experiments (Fig. 3) have technical specifications similar to the original CODE drifters designed by Dr. Russ Davis of the Scripps Institution of Oceanography. Each drifter consists of a thin (15cm in diameter) 1m long cylindrical metal body with 0.5m long metal foldable cross-shaped upper arms and lower legs that held a rectangular cloth sail. The drifter is attached by four 20cm rope segments to 4 small (15cm in diameter) round plastic surface buoys for floatation. The round buoy shape minimizes the wave effects compared to flat buoys, the ropes minimizes tilting of the sail compared to the “solid-neck” drifter design, and the large body-to-buoy size ratio insures good water-following characteristics. The drifters are equipped with GPS transmitters that provide positioning fixes every 5min. Based on land tests conducted prior to deployment, the STD of the GPS positioning error is on the order of a few meters (exact values depend on the sky view and location). Estimates of the expected wind slippage of CODE type drifters with standard sails such as ours are 1–2 cm/s in light wind conditions similar to those during our field experiment \cite{Ohlmann2007,Poulain2009}. 
Drifters of the same design are routinely used by the U.S. Coast Guard in SAR operations, as well as in our previous field experiments south of Martha’s Vineyard, MA \cite{rypina2014eulerian,rypina2016Invest}. 
%
%
\subsection{Manikins}\label{App:Manikins}
We used OSCAR Water Rescue Training manikins (Fig. 3) manufactured by Emerald Marine Products (Edmonds, WA) for man-overboard rescue training. Each manikin consists of eight heavy-duty vinyl body parts, PVC fill/drain fittings, six stainless steel joints and two galvanized lifting shackles. The manikin filled with water replicates an 82 kg rescue victim, 1.83 m tall and 0.46 m wide (chest). For an accurate simulation of a person in water, the manikin is filled with water to float at chest level. The manikins are equipped with the same GPS transmitters used for drifters, which provide positioning fixes every 5min.
%

\bibliographystyle{ieeetr}
\bibliography{/Users/serram/Dropbox/ScientificLiterature/BibFiles/ReferenceList3}

\begin{thebibliography}{10}

\bibitem{DeathAtSeaMediterr2017}
E.~Steinhilper and R.~Gruijters, ``Border deaths in the mediterranean: What we
  can learn from the latest data.'' (2017),
  https://www.law.ox.ac.uk/research-subject-groups/centre-criminology/centreborder-criminologies/blog/2017/03/institutional.

\bibitem{DeathAtSeaStatsMSF2017}
J.~H. o. M. i.~L. Pagotto, M.~Search, and Rescue, ``How to stop the rising tide
  of death in mediterranean.'' (2017),
  https://www.msf.org/migration-how-stop-rising-tide-death-mediterranean.

\bibitem{UNHCR_StrenghtenSARMediterr2018}
C.~Yaxley, ``As mediterranean sea arrivals decline and death rates rise, unhcr
  calls for strengthening of search and rescue.'' (2018),
  https://www.unhcr.org/afr/news/briefing/2018/7/5b3f270a4/mediterranean-sea-arrivals-decline-death-rates-rise-unhcr-calls-strengthening.html.

\bibitem{Breivik2013}
{\O}.~Breivik, A.~Allen, C.~Maisondieu, and M.~Olagnon, ``Advances in search
  and rescue at sea,'' 2013.

\bibitem{stone2013search}
L.~Stone, ``Search theory,'' {\em Encyclopedia of Operations Research and
  Management Science}, pp.~1366--1378, 2013.

\bibitem{allen1996performance}
A.~Allen, ``{Performance of GPS/Argos self-locating datum marker buoys
  (SLDMBs}),'' in {\em OCEANS'96. MTS/IEEE. Prospects for the 21st Century.
  Conference Proceedings}, vol.~2, pp.~857--861, IEEE, 1996.

\bibitem{Kratzke2010}
T.~Kratzke, L.~Stone, and J.~Frost, ``{Search and rescue optimal planning
  system},'' in {\em Information Fusion (FUSION), 2010 13th Conference on},
  pp.~1--8, IEEE, 2010.

\bibitem{SerraHaller2015}
M.~Serra and G.~Haller, ``{Objective Eulerian coherent structures},'' {\em
  Chaos}, vol.~26, no.~5, p.~053110, 2016.

\bibitem{LCSHallerAnnRev2015}
G.~Haller, ``{Lagrangian coherent structures},'' {\em Annual Rev. Fluid. Mech},
  vol.~47, pp.~137--162, 2015.

\bibitem{TruesdellNoll2004}
C.~Truesdell and W.~Noll, {\em {The non-linear field theories of mechanics}}.
\newblock Springer, 2004.

\bibitem{Kirincich2012}
A.~Kirincich, T.~De~Paolo, and E.~Terrill, ``{Improving HF radar estimates of
  surface currents using signal quality metrics, with application to the MVCO
  high-resolution radar system},'' {\em J. Atmos. Oceanic Technol.}, vol.~29,
  no.~9, pp.~1377--1390, 2012.

\bibitem{richardson2009drifters}
P.~Richardson, ``Drifters and floats,'' {\em {Elements of Physical
  Oceanography: A derivative of the Encyclopedia of Ocean Sciences}}, p.~89,
  2009.

\bibitem{DAsaro2018}
E.~DAsaro, A.~Shcherbina, J.~Klymak, J.~Molemaker, G.~Novelli, C.~Guigand,
  A.~Haza, B.~Haus, E.~Ryan, G.~Jacobs, {\em et~al.}, ``{Ocean convergence and
  the dispersion of flotsam},'' {\em Proceedings of the National Academy of
  Sciences}, vol.~115, no.~6, pp.~1162--1167, 2018.

\bibitem{maio2016evaluation}
A.~Di~Maio, M.~Martin, and R.~Sorgente, ``{Evaluation of the search and rescue
  LEEWAY model in the Tyrrhenian Sea: a new point of view},'' {\em Natural
  Hazards and Earth System Sciences}, vol.~16, no.~8, pp.~1979--1997, 2016.

\bibitem{Bellomo2015}
L.~Bellomo, A.~Griffa, S.~Cosoli, P.~Falco, R.~Gerin, I.~Iermano,
  A.~Kalampokis, Z.~Kokkini, A.~Lana, M.~Magaldi, {\em et~al.}, ``{Toward an
  integrated HF radar network in the Mediterranean Sea to improve search and
  rescue and oil spill response: the TOSCA project experience},'' {\em Journal
  of Operational Oceanography}, vol.~8, no.~2, pp.~95--107, 2015.

\bibitem{Haley2010}
P.~Haley and P.~Lermusiaux, ``{Multiscale two-way embedding schemes for
  free-surface primitive equations in the Multidisciplinary Simulation,
  Estimation and Assimilation System},'' {\em Ocean Dyn}, vol.~60, no.~6,
  pp.~1497--1537, 2010.

\bibitem{lermusiaux_JCP2006}
P.~F.~J. {Lermusiaux}, ``{Uncertainty estimation and prediction for
  interdisciplinary ocean dynamics},'' {\em Journal of Computational Physics},
  vol.~217, no.~1, pp.~176--199, 2006.

\bibitem{lermusiaux_et_al_O2006b}
P.~F.~J. {Lermusiaux}, C.-S. {Chiu}, G.~G. {Gawarkiewicz}, P.~{Abbot}, A.~R.
  {Robinson}, R.~N. {Miller}, P.~J. {Haley}, Jr, W.~G. {Leslie}, S.~J.
  {Majumdar}, A.~{Pang}, and F.~{Lekien}, ``Quantifying uncertainties in ocean
  predictions,'' {\em Oceanography}, vol.~19, no.~1, pp.~92--105, 2006.

\bibitem{ArnoldODE1973}
V.~Arnold, {\em Ordinary Differential Equations}.
\newblock MIT Press, Boston, 1973.

\bibitem{GuckenheimerHolmes1983}
J.~Guckenheimer and P.~Holmes, {\em Nonlinear oscillations, dynamical systems,
  and bifurcations of vector fields}, vol.~42.
\newblock Springer Science \& Business Media, 1983.

\bibitem{Beronetal08b}
F.~Beron-Vera, M.~Olascoaga, and G.~Goni, ``{Oceanic mesoscale vortices as
  revealed by Lagrangian coherent structures},'' {\em Geophys. Res. Lett.},
  vol.~35, p.~L12603, 2008.

\bibitem{Olascoaga2013}
M.~Olascoaga, F.~Beron-Vera, G.~Haller, J.~Trinanes, M.~Iskandarani, E.~Coelho,
  B.~Haus, H.~Huntley, G.~Jacobs, A.~Kirwan, {\em et~al.}, ``{Drifter motion in
  the Gulf of Mexico constrained by altimetric Lagrangian coherent
  structures},'' {\em Geophys. Res. Lett.}, vol.~40, no.~23, pp.~6171--6175,
  2013.

\bibitem{kirincich2016remote}
A.~Kirincich, ``{Remote sensing of the surface wind field over the coastal
  ocean via direct calibration of HF radar backscatter power},'' {\em J.
  Atmosph. and Oceanic Tech.}, vol.~33, no.~7, pp.~1377--1392, 2016.

\bibitem{Kirincich2016}
A.~Kirincich, ``{The occurrence, drivers, and implications of submesoscale
  eddies on the Martha's Vineyard inner shelf},'' {\em J. Phys. Oceanogr.},
  vol.~46, no.~9, pp.~2645--2662, 2016.

\bibitem{hadjighasem2017critical}
A.~Hadjighasem, M.~Farazmand, D.~Blazevski, G.~Froyland, and G.~Haller, ``{A
  critical comparison of Lagrangian methods for coherent structure
  detection},'' {\em Chaos}, vol.~27, no.~5, p.~053104, 2017.

\bibitem{haley_et_al_OM2015}
P.~J. Haley, Jr., A.~Agarwal, and P.~F.~J. Lermusiaux, ``Optimizing velocities
  and transports for complex coastal regions and archipelagos,'' {\em Ocean
  Modeling}, vol.~89, pp.~1--28, 2015.

\bibitem{logutov_lermusiaux_OM2008}
O.~G. {Logutov} and P.~F.~J. {Lermusiaux}, ``Inverse barotropic tidal
  estimation for regional ocean applications,'' {\em Ocean Modelling}, vol.~25,
  no.~1--2, pp.~17--34, 2008.

\bibitem{lermusiaux_JAOT2002}
P.~F.~J. Lermusiaux, ``On the mapping of multivariate geophysical fields:
  Sensitivities to size, scales, and dynamics,'' {\em Journal of Atmospheric
  and Oceanic Technology}, vol.~19, no.~10, pp.~1602--1637, 2002.

\bibitem{Ohlmann2007}
C.~Ohlmann, P.~White, L.~Washburn, B.~Emery, E.~Terrill, and M.~Otero,
  ``{Interpretation of coastal HF radar--derived surface currents with
  high-resolution drifter data},'' {\em J. Atmos. Oceanic Technol.}, vol.~24,
  no.~4, pp.~666--680, 2007.

\bibitem{Poulain2009}
R.~Poulain, P.and~Gerin, E.~Mauri, and R.~Pennel, ``{Wind effects on drogued
  and undrogued drifters in the eastern Mediterranean},'' {\em J. Atmos.
  Oceanic Technol.}, vol.~26, no.~6, pp.~1144--1156, 2009.

\bibitem{rypina2014eulerian}
I.~Rypina, A.~Kirincich, and I.~Limeburner, R.and~Udovydchenkov, ``{Eulerian
  and Lagrangian correspondence of high-frequency radar and surface drifter
  data: Effects of radar resolution and flow components},'' {\em J. Atm. and
  Oceanic Tech.}, vol.~31, no.~4, pp.~945--966, 2014.

\bibitem{rypina2016Invest}
I.~Rypina, A.~Kirincich, S.~Lentz, and M.~Sundermeyer, ``{Investigating the
  eddy diffusivity concept in the coastal ocean},'' {\em J. Phys. Oceangr.},
  vol.~46, no.~7, pp.~2201--2218, 2016.

\end{thebibliography}

\end{document}